\documentclass[superscriptaddress,nobibnotes,amsmath,amssymb,notitlepage,twocolumn,prl]{revtex4-1}

\usepackage[colorlinks=true,linkcolor=blue,citecolor=blue,urlcolor=blue]{hyperref}

\usepackage{setspace} 
\usepackage{graphicx} \graphicspath{{./img/}} 
\usepackage{amsmath}
\usepackage{color}
\usepackage{amsmath}
\usepackage{amssymb}
\usepackage{verbatim}
\usepackage{latexsym}
\usepackage{enumerate} 
\usepackage{bm} 
\usepackage{ulem} 

\newcommand{\Te}{T_{\rm e}}

\newcommand*{\heading}[1]{\belowpdfbookmark{#1}{#1}{\textit{#1.---}}\ignorespaces}
\let\section\heading 

\begin{document}

\title{
Photo-induced electronic and spin topological phase transitions in monolayer bismuth
}



\newcommand{\TCM}{Theory of Condensed Matter Group, Cavendish Laboratory, University of Cambridge, J.\,J.\,Thomson Avenue, Cambridge CB3 0HE, UK}
\newcommand{\Materials}{Department of Materials Science and Metallurgy, University of Cambridge, 27 Charles Babbage Road, Cambridge CB3 0FS, UK}
\newcommand{\HarvardSeas}{John A.~Paulson School of Engineering and Applied Sciences, Harvard University, Cambridge, Massachusetts 02138, USA}


\author{Bo Peng}
\email{bp432@cam.ac.uk}
\affiliation{\TCM}

\author{Gunnar F. Lange}
\affiliation{\TCM}

\author{Daniel Bennett}
\affiliation{\HarvardSeas}

\author{Kang Wang} 
\affiliation{\Materials}

\author{Robert-Jan Slager}
\affiliation{\TCM}

\author{Bartomeu Monserrat} 
\email{bm418@cam.ac.uk}
\affiliation{\TCM}
\affiliation{\Materials}

\date{\today}

\begin{abstract}
Ultrathin bismuth exhibits rich physics including strong spin-orbit coupling, ferroelectricity, nontrivial topology, and light-induced structural dynamics. We use {\it ab initio} calculations to show that light can induce structural transitions to four transient phases in bismuth monolayers. These light-induced phases exhibit nontrivial topological character, which we illustrate using the recently introduced concept of spin bands and spin-resolved Wilson loops. Specifically, we find that the topology changes via the closing of the electron and spin band gaps during photo-induced structural phase transitions, leading to distinct edge states. Our study provides strategies to tailor electronic and spin topology via ultrafast control of photo-excited carriers and associated structural dynamics.
\end{abstract}

\maketitle

Bismuth exhibits a rich spectrum of topological phases in all its forms, from monolayer/bilayer to thin film and to bulk\,\cite{Murakami2006,Murakami2007,Fu2007,Murakami2011a,Wada2011,Khomitsky2014,Ning2014,Drozdov2014,Lu2015,Ito2016,Bian2016a,Schindler2018,Nayak2019,Hsu2019,Koenig2021,Bai2022}. Strong spin-orbit coupling (SOC) motivated early studies of the boundary states of bismuth in the context of spin splitting\,\cite{Koroteev2004,Pascual2004,Sugawara2006,Hirahara2006,Hirahara2007,Takayama2011,Xiao2012a,Takayama2015}. With the advent of topological materials, bismuth was one of the first proposed quantum spin Hall (QSH) insulators in 2D\,\cite{Murakami2006,Murakami2007} and topological insulators in 3D\,\cite{Fu2007}, both of which have been confirmed experimentally\,\cite{Ning2014,Drozdov2014,Lu2015,Ito2016}. More recently, it has been experimentally shown that bismuth exhibits exotic properties such as topological crystalline insulator states and higher-order topology\,\cite{Schindler2018,Nayak2019,Hsu2019}. 


In its ultrathin form, bismuth exhibits multiple structural phases, predicted theoretically\,\cite{Singh2019} and characterized experimentally\,\cite{Nagao2004}. It has been predicted that a puckered $Pmn2_1$ monolayer phase exhibits elemental ferroelectricity with switchable in-plane polarization, and this phase can be understood with respect to a paraelectric $Pmna$ phase with no puckering\,\cite{Xiao2018}. Recently, such ferroelectric switching has been confirmed experimentally using an in-plane electric field produced by scanning probe microscopy\,\cite{Gou2023}, holding promise for various applications in memory\,\cite{Hong2023} and energy harvesting\,\cite{Qian2023} devices. Additionally, several experimental and theoretical works have proposed distinct topological properties for both phases, including Dirac\,\cite{Kowalczyk2020} and Weyl states\,\cite{Lu2023} for the ferroelectric $Pmn2_1$ phase, 2D topological insulator states\,\cite{Lu2015} for the paraelectric $Pmna$ phase, and generalized QSH states for both phases\,\cite{Bai2022,Wang2022b}. Overall, monolayer bismuth promises to be a fruitful platform to study the interplay between structural, topological, and ferroelectric/paraelectric phases, especially how the topological phase evolves as the structural phase transition takes place, as found in other layered materials\,\cite{Ezawa2013,Mocatti2023,Sie2019,Zhang2019b,McIver2020}.





Among the external stimuli typically used to tune structural phases of bismuth, light can be a relatively simple and economical option. The structural properties of bismuth are sensitive to laser energy and fluence, and even small changes in electronic occupation significantly affect its potential energy surface\,\cite{Faure2013}. Bulk bismuth has served as a reference material for studying photo-excited carrier and structural dynamics for decades\,\cite{Hase2002,Murray2005,Murray2007,Fritz2007,Johnson2009,Timrov2012,Papalazarou2012,Faure2013,Johnson2013,Perfetti2015,Murray2015,OMahony2019}. 
In this context, we ask whether monolayer bismuth may be a good candidate for realizing photo-switchable 
topological properties through phase transitions tuned by photoexcitation. To this end, it would be insightful to investigate whether light can induce transitions between the topologically distinct ferroelectric $Pmn2_1$ and paraelectric $Pmna$ phases, or lead to other hidden phases that cannot be accessed in the dark.

In this Letter, we show that light can be used to control both structural and topological properties of ferroelectric and paraelectric bismuth monolayers. Varying the photo-excited carrier density and electronic temperature, four previously unreported transient inversion symmetric phases are stabilized. We perform a spin band structure analysis\,\cite{Prodan2009,Lin2022,Lange2023} to characterize the topological properties and find that the two dark phases exhibit different QSH insulating behavior with different spin Chern numbers. Remarkably, photo-induced phase transitions drive the ferroelectric and paraelectric phases into 2D Dirac and nodal-line semimetals respectively, each with distinct topological edge states.

\begin{figure*}
\centering
\includegraphics[width=\textwidth]{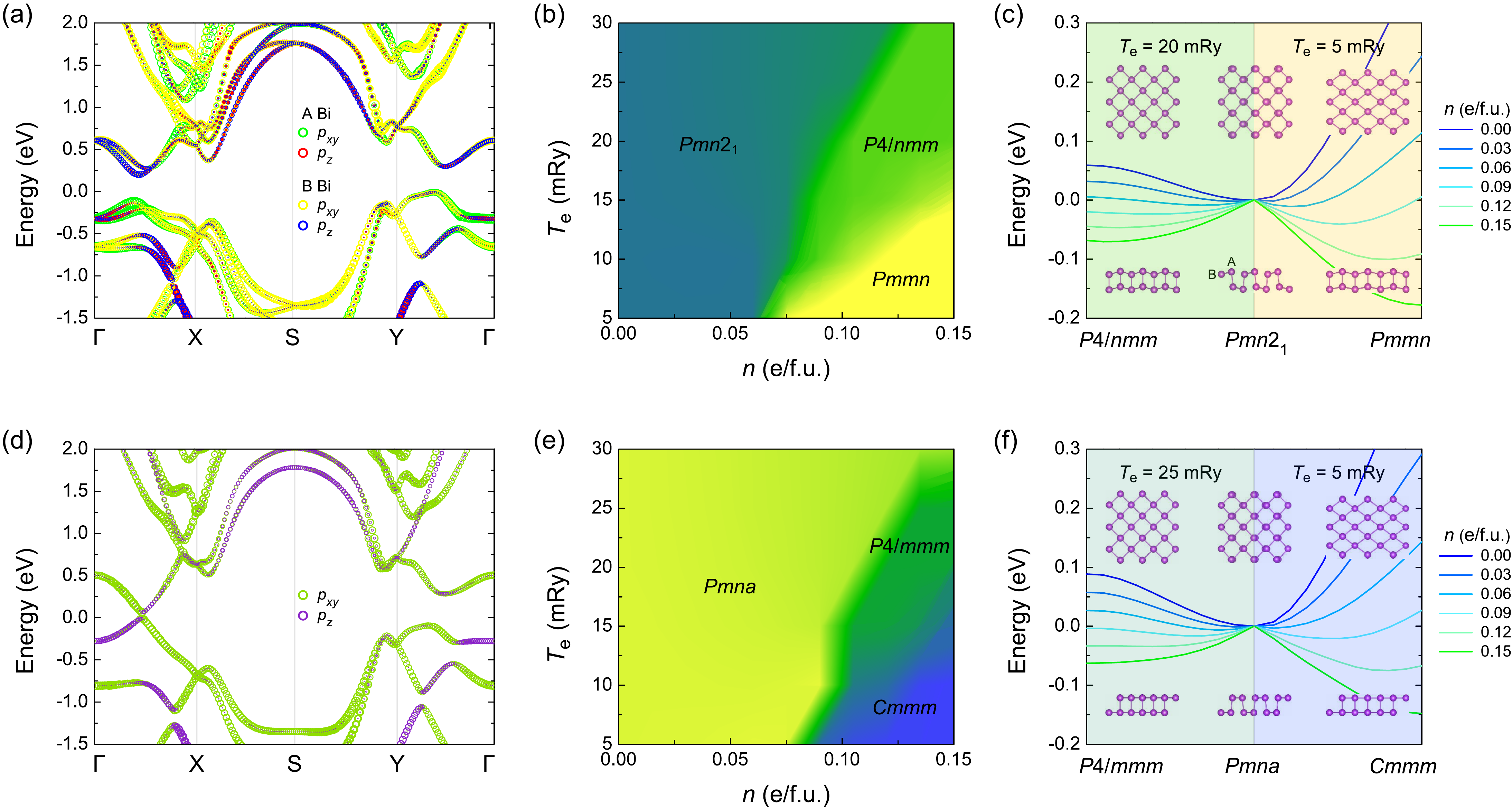}
\caption{
(a) Orbital-projected band structure of $Pmn2_1$ Bi. 
(b) Phase diagram for $Pmn2_1$ Bi as a function of $n$ and $\Te$. 
(c) Potential energy surfaces (per unit cell) along the transition paths $Pmn2_1 \rightarrow P4/nmm$ and $Pmn2_1 \rightarrow Pmmn$ at the ground and excited states.
(d) Orbital-projected band structure of $Pmna$ Bi. 
(e) Phase diagram for $Pmna$ Bi as a function of $n$ and $\Te$. 
(f) Potential energy surfaces (per unit cell) along the transition paths $Pmna \rightarrow P4/mmm$ and $Pmna \rightarrow Cmmm$ at the ground and excited states. The crystal structure of each phase is sketched in (c) and (f).
}
\label{phases} 
\end{figure*}

Photoexcitation is simulated using two chemical potentials for thermalized electrons and holes respectively\,\cite{Tangney1999,Tangney2002,Murray2007,Paillard2016,Haleoot2017,Paillard2017,Paillard2019,Gu2021,Marini2021,Marini2021a,Marini2022}, constraining a fixed number of electrons to the conduction bands and holes to the valence bands\,\cite{Peng2020,Peng2020a,Peng2022d}. Compared to the single Fermi-Dirac distribution\,\cite{Zijlstra2006,Johnson2008,Zijlstra2010,Giret2011}, the setup of two Fermi-Dirac distributions describes an electron-hole plasma state (where excitons are completely screened) that provides an accurate description of optically excited structural dynamics in semimetallic systems such as bulk bismuth\,\cite{Murray2005,Murray2007,Fritz2007,OMahony2019}, and is suitable for studying bismuth monolayers with a well-defined band gap. Density functional theory calculations are performed using the {\sc Quantum Espresso} package\,\cite{Giannozzi2009,Giannozzi2020} with the structural properties cross checked using {\sc Abinit} \,\cite{gonze2016,gonze2020}. 
A fully-relativistic projector augmented wave pseudopotential\,\cite{DalCorso2012} is used with the PBEsol 
functional\,\cite{Perdew2008}. Kinetic energy cutoffs of $120$\,Ry and $500$\,Ry are chosen for the wavefunction and charge density respectively, with a 
{$\boldsymbol{k}$-point} grid of $12\times 12\times 1$. 
A vacuum spacing of {24 \AA}\ is employed in combination with the truncation of the Coulomb interaction in the $z$ 
direction\,\cite{Sohier2017a}.
To analyze the topological properties, maximally localized Wannier functions\,\cite{Marzari1997,Souza2001,Marzari2012} for the $6p$ orbitals are generated using {\sc wannier90}\,\cite{Mostofi2008,Mostofi2014,Pizzi2020}. 
The edge states are computed using {\sc WannierTools}\,\cite{Wu2018}.

The electronic structure of the ferroelectric $Pmn2_1$ phase is shown in Fig.\,\ref{phases}(a). Performing a structural relaxation for different photo-excited carrier densities $n$ and electronic temperatures $\Te$, two paraelectric phases can be obtained [Fig.\,\ref{phases}(b)]. This is similar to recent studies of ferroelectric perovskites, where the paraelectric phase was predicted to be stabilized by light as carriers effectively screen the polarization\,\cite{Paillard2017,Peng2020a}. At $\Te = 5$\,mRy ($\sim 790$\,K), the $Pmmn$ phase is stabilized at {$n=0.075$\,e/f.u.} (electron per formula unit). Increasing $\Te$ and $n$, the $P4/nmm$ phase becomes energetically favorable.
The crystal structures of the three phases are sketched in the inset of Fig.\,\ref{phases}(c). The buckling of the Bi atoms in the same sublayer (labeled A and B) in the $Pmn2_1$ phase is also observed in the photo-induced phases. However, the atoms in the sublayers become vertically aligned, restoring inversion symmetry and preventing any polarization. Although both photo-induced phases are paraelectric, they are distinguished by their rectangular ($Pmmn$) and square ($P4/nmm$) lattices.

\begin{figure*}
\centering
\includegraphics[width=\textwidth]{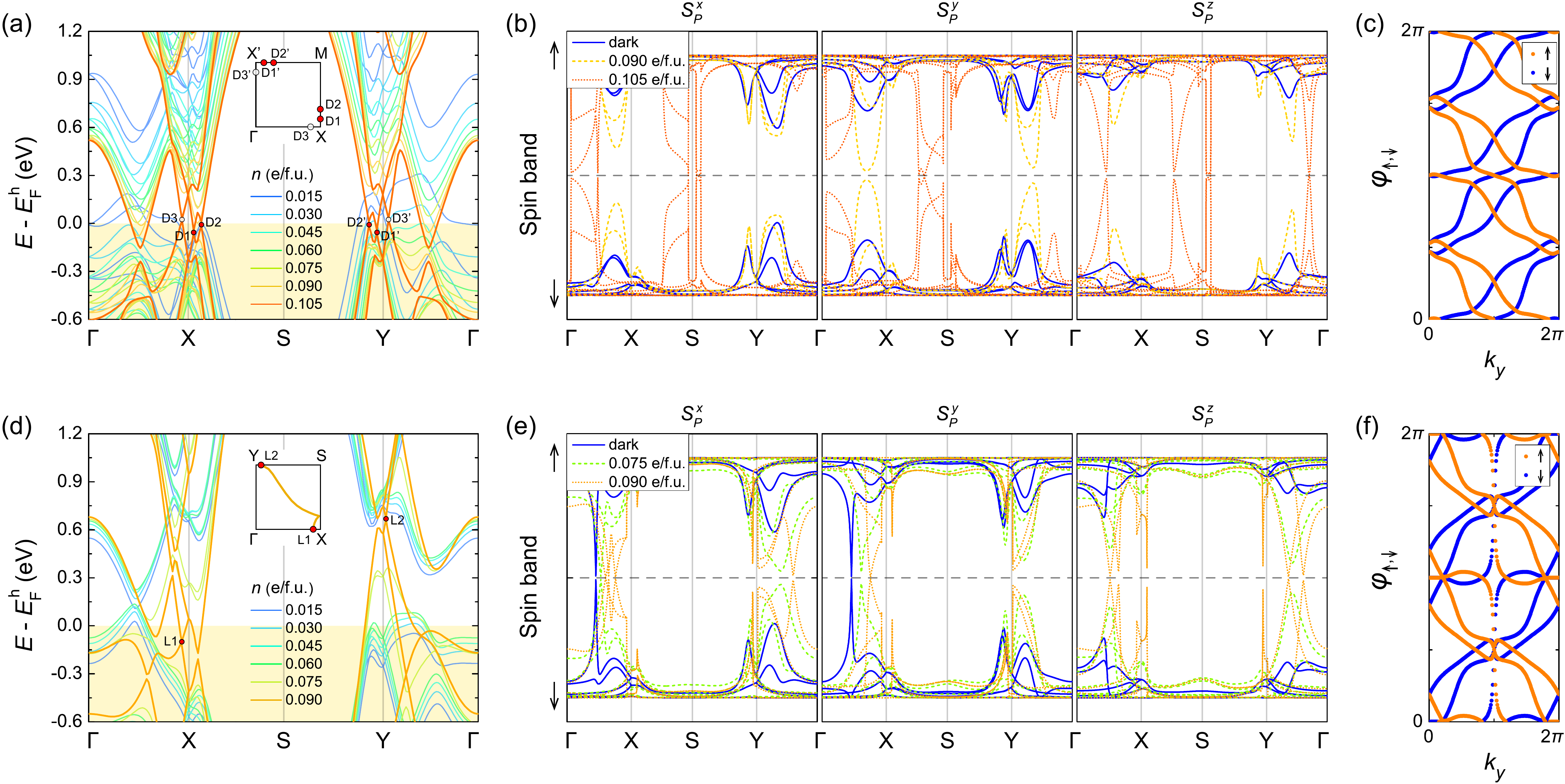}
\caption{
(a) Band structure of $Pmn2_1$ Bi as a function of $n$ for $\Te = 20$\,mRy, with the quasi-Fermi level for holes $E_{\textrm{F}}^{\textrm{h}}$ set to zero. At $n = 0.105$\,e/f.u., the system becomes a 2D Dirac semimetal with space group $P4/nmm$. The inset shows the 2D Brillouin zone (BZ) with the Dirac points denoted as D. 
(b) Spin band structures $S_P^{x,y,z}$ of $Pmn2_1$ Bi at $n$ = 0, 0.090, $0.105$\,e/f.u. for $\Te = 20$\,mRy. 
(c) Spin Wilson loop $\varphi_{\uparrow,\downarrow}$ of $Pmn2_1$ Bi in the dark, displaying a winding of $2$ in each spin channel.
(d) Band structures of $Pmna$ Bi as a function of $n$ for $\Te = 5$\,mRy, with  $E_{\textrm{F}}^{\textrm{h}}$ set to zero. At $n = 0.090$\,e/f.u., the system becomes a 2D nodal-line semimetal with space group $Cmmm$. The inset shows the BZ, with the nodal line. 
(e) Spin band structures $S_P^{x,y,z}$ of $Pmna$ Bi at $n$ = 0, 0.15, $0.090$\,e/f.u. for $\Te = 5$\,mRy. 
(f) Spin Wilson loop $\varphi_{\uparrow,\downarrow}$ of $Pmna$ Bi in the dark, displaying a winding of $3$ for each spin channel.
}
\label{band} 
\end{figure*}

To understand whether the photo-induced phases are metastable or transient, we compute the potential energy surfaces along the transition paths from $Pmn2_1$ ($n=0$\,e/f.u.) to $P4/nmm$ ($n=0.105$\,e/f.u., $\Te = 20$\,mRy) and $Pmmn$ ($n=0.150$\,e/f.u., $\Te = 5$\,mRy), respectively, for several fixed values of $n$ [Fig.\,\ref{phases}(c)]. 
In the dark, the $P4/nmm$ phase is unstable. 
Increasing $n$, the $P4/nmm$ phase becomes stable while the $Pmn2_1$ phase becomes unstable, indicating that the $P4/nmm$ phase is transient and can only be accessed by photoexcitation. Similar behavior is observed for the transition $Pmn2_1 \rightarrow Pmmn$ at $\Te = 5$\,mRy, (for the transition path $P4/nmm$ $\rightarrow$ $Pmmn$ with decreasing $\Te$, see the Supplemental Material\,\cite{PRL2023SM_Bi}).

We also consider photo-induced transitions starting from the paraelectric $Pmna$ phase, which is energetically less favorable than the ferroelectric $Pmn2_1$ phase by 16\,meV per unit cell in the dark within the static lattice approximation, in agreement with experimental reports (for the relative stability of the $Pmn2_1$ and $Pmna$ phases, see the Supplemental Material\,\cite{PRL2023SM_Bi}). 
We note that the paraelectric $Pmna$ phase has been obtained experimentally through the application of an in-plane electric field\,\cite{Gou2023} or at room temperature on a substrate\,\cite{Kowalczyk2020}. Our calculations also show that the paraelectric phase becomes stable at room temperature, in agreement with experiment, and we also predict that it becomes stable by the application of strain, but we find that the phase transition $Pmn2_1 \to Pmna$ cannot be driven using light (details in the Supplemental Material\,\cite{PRL2023SM_Bi}).

Unlike the $Pmn2_1$ phase, all four atoms in the $Pmna$ phase contribute equally to the electronic states near the Fermi level [Fig.\,\ref{phases}(d)]. As a result, photoexcitation cannot make the two Bi atoms in the same sublayer asymmetric, i.e.~puckering cannot be induced. Two phases are realized upon photoexcitation of the $Pmna$ phase: $Cmmm$ at low $\Te$ and $P4/mmm$ at high $\Te$ [Fig.\,\ref{phases}(e)]. Neither phase exhibits buckling [Fig.\,\ref{phases}(f)], contrary to the phases in Fig.\,\ref{phases}(b). The potential energy surfaces from the $Pmna$ phase in the dark to the photo-excited 
phases are shown in Fig.\,\ref{phases}(f), indicating that they are also transient and cannot be realized in the dark. 

\begin{figure*}
\centering
\includegraphics[width=\textwidth]{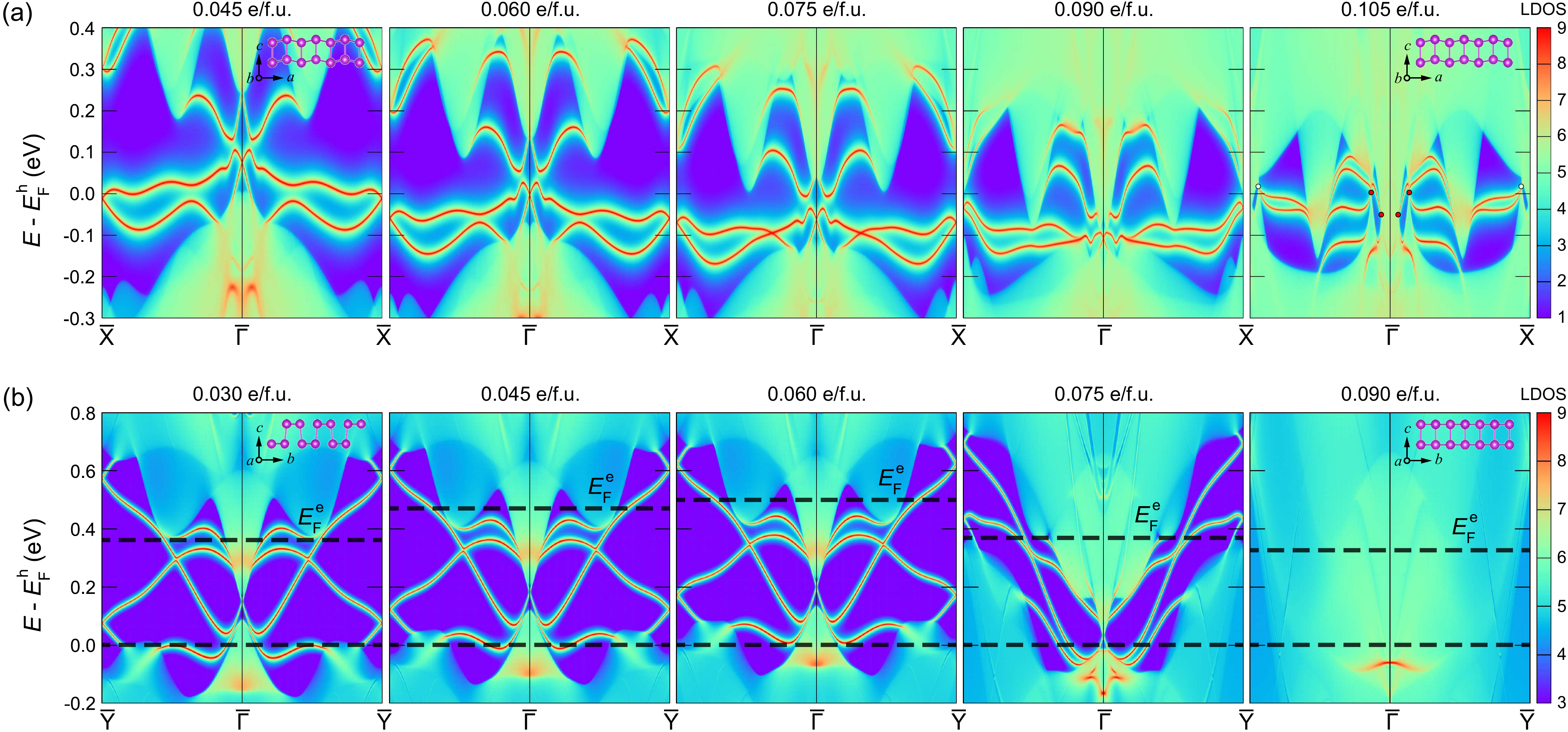}
\caption{
(a) $(010)$ edge states of $Pmn2_1$ Bi at $n = 0.045-0.105$\,e/f.u. for $\Te = 20$\,mRy. The transition to $P4/nmm$ occurs at $0.105$\,e/f.u.
(b) $(100)$ edge states of $Pmna$ Bi at $n = 0.030-0.090$\,e/f.u. for $\Te = 5$\,mRy. The transition to $Cmmm$ occurs at $0.090$\,e/f.u.
}
\label{surface} 
\end{figure*}

Figure\,\ref{band}(a) shows the band structure as a function of $n$ starting from the $Pmn2_1$ phase. Upon photoexcitation, the gap decreases until the phase transition to $P4/nmm$ at $n=0.105$\,e/f.u., where it closes, indicating that a topological phase transition accompanies the structural transition. 
For a more refined diagnosis of the topological properties with fully relativistic SOC included, we consider the spin degree of freedom and examine the spin band structures\,\cite{Prodan2009,Lin2022,Lange2023}. 
We consider the projected spin operator along axis $\hat{\boldsymbol{n}}$, ${S_P^{\hat{\boldsymbol{n}}}(\boldsymbol{k}) = P(\boldsymbol{k})S^{\hat{\boldsymbol{n}}}P(\boldsymbol{k})}$, where ${S^{\hat{\boldsymbol{n}}} = \mathbf{1}_{\mathrm{orb}}\otimes \hat{\boldsymbol{n}}\cdot \boldsymbol{\sigma}}$ and  $P(\boldsymbol{k})$ is the projector onto occupied bands. Importantly, this operator is gauge-invariant provided the electronic band gap does not close\,\cite{Lin2022,Lange2023}. Diagonalization of $S_P^{\hat{\boldsymbol{n}}}$ leads to spin bands along $\hat{\boldsymbol{n}}$, which generically do not coincide with the electronic bands and which, for sufficiently strong SOC, can be gapless even when the electronic bands are gapped. Whenever the spin up bands remain separated from the spin down bands however (referred to as the spin gap being open), the spin up/down sectors along $\hat{\boldsymbol{n}}$ are well-defined and their topology can be studied separately, leading to a generalized, $\mathbb{Z}$-valued QSH effect\,\cite{Prodan2009,Prodan2010,Yang2011,Bai2022}, which relies on the presence of a gap in both the energy and the spin spectrum. 
The spin band structures 
of the $Pmn2_1$ phase are shown in Fig.\,\ref{band}(b). In the dark, the electronic band gap is open, 
and the spin gap is also open along $x,y,z$, implying that the spin topology is well defined along these directions. By considering the spin-$z$ resolved Wilson loop winding\,\cite{Bouhon2019,Lin2022,Lange2023} in Fig.\,\ref{band}(c), we find a spin Chern number $C_S=-2$. 
Thus, the $Pmn2_1$ phase in the dark is an exotic doubled QSH insulator with trivial Kane-Mele invariant $z_2=C_S\ \mathrm{mod}\ 2$\,\cite{Kane2005,Kane2005a,Prodan2009,Lin2022}, in agreement with Ref.\,\cite{Bai2022}. 

The photo-induced $P4/nmm$ phase is a 2D Dirac semimetal because inversion symmetry is restored\,\cite{Young2015,Meng2022,Ding2022,Yu2022a}. The Dirac points D1 and D2 [marked in red in Fig.\,\ref{band}(a)] appear along the X--S high-symmetry line and remain nearly fixed for larger $n$, while another Dirac point D3 (marked in white), emerges along $\Gamma$--X for $n>0.105$\,e/f.u.
Note that the spin band structure in Fig.\,\ref{band}(b) ceases to be gauge-invariant when the electronic band gap closes. We conclude that the doubled QSH insulator becomes a 2D Dirac semimetal when the structural phase transition takes place. 

The band structure as a function of $n$ starting from the $Pmna$ phase is shown in Fig.\,\ref{band}(d). Under weak or vanishing photoexcitation, the electronic band gap remains open while the spin gap closes for the in-plane spin components $S_P^x$ and $S_P^y$, implying that the spin topology is not well defined along the in-plane directions [Fig.\,\ref{band}(e)]. For the out-of-plane spin component $S_P^z$, the spin gap remains open as long as the electronic gap remains open, suggesting that the effect of SOC is weaker in this direction. 
For the $z$-directed spin, we find $C_S=-3$ in Fig.\,\ref{band}(f). This suggests that the $Pmna$ phase is a QSH insulator as {$z_2=C_S$ mod $2=1$}, in agreement with Ref.\,\cite{Wang2022b}. However, by considering the projected spin band structure within fully relativistic SOC, we can identify this phase as an exotic tripled QSH insulator as long as the $S_P^z$ gap remains open. The band structures for photo-excited $Pmna$ Bi are shown in Fig.\,\ref{band}(d). At $n = 0.090$\,e/f.u., the phase transition to $Cmmm$ occurs, and the two bands become degenerate along a nodal line. Thus, the light-induced structural and topological phase transitions occur simultaneously from a tripled QSH insulator 
to a 2D Dirac nodal-line semimetal.

To investigate the boundary signatures of the spin topological phases\,\cite{Hasan2010,Qi2011,Rhim2018}, we study the electronic edge states. For the $Pmn2_1$ phase, we plot the $(010)$ edge states in Fig.\,\ref{surface}(a), from 0.045 to 0.105\,e/f.u. for $\Te$ = 20\,mRy. Before the phase transition, the doubled QSH insulator displays two pairs of helical edge states which are doubly degenerate at $\overline{\rm X}$ and $\overline{\Gamma}$ 
(the upper pair partially overlaps with the bulk). In a doubled QSH insulator with conserved $s_z$, i.e.~with quantized $S^{z}_P$ eigenvalues, we expect two helical edge states crossing the gap. As $s_z$ is not conserved in our system, the edge states can elastically backscatter pairwise, opening a gap between them\,\cite{Kane2005}, as shown in Fig.\,\ref{surface}(a). These edge states are remnants of the bulk topology, but they are not topologically protected, in agreement with the trivial $z_2$ invariant. During the phase transition to the Dirac semimetal, the helical edge states merge into the projections of the bulk Dirac points, connecting the projected Dirac points. 

In Fig.\,\ref{surface}(b) we show the $(100)$ edge states of the paraelectric $Pmna$ phase from $0.06$ to $0.090$\,e/f.u. for $\Te = 5$\,mRy. We see a pair of helical edge states crossing the gap, in agreement with the usual $z_2=1$ phases\,\cite{Kane2005,Kane2005a}. Interestingly, however, we also observe additional edge states, which are weakly gapped out and appear robust up to the phase transition point. Similar to the $Pmn2_1$ phase, these are likely remnant edge states, arising due to the fact that the bulk has a high spin Chern number without $s_z$ conservation. After the phase transition, the edge states of the Dirac nodal-line semimetal are hidden in the projections of the bulk states and are only visible around the $\overline{\Gamma}$ point at about $-0.1$\,eV. The split of the edge bands away from the $\overline{\Gamma}$ point has been experimentally observed in Sb$(111)$ and Bi$(111)$ thin films\,\cite{Koroteev2004,Pascual2004,Sugawara2006,Hirahara2006,Hirahara2007,Takayama2011,Xiao2012a,Takayama2015}, and has been attributed to the strong SOC\,\cite{Hofmann2006,Bihlmayer2007} (for edge oxidation, see the Supplemental Material\,\cite{PRL2023SM_Bi}).


For experimental observation, the topological properties of 2D Bi\,\cite{Kowalczyk2020} and Sb\,\cite{Lu2021,Lu2022} have been investigated intensively. In terms of photoexcitation, 
photo-induced structural dynamics in Bi have been well studied by time-resolved X-ray diffraction measurements\,\cite{Fritz2007,Johnson2013}, while time-resolved angle-resolved photoemission spectroscopy (trARPES) has been widely used to visualize the electronic structure evolution of photo-excited bismuth and black phosphorus\,\cite{Faure2013,Zhou2023}. The photoexcitation energy (1.6\,eV) and fluence ($0.2-3$\,mJ/cm$^2$) used in these experiments\,\cite{Faure2013,Zhou2023} can readily generate photo-excited carrier densities $n > 0.15$\,e/f.u., exceeding the maximum carrier density used in this study to observe all the predicted phenomena (for the maximum number of electron-hole pairs, see the Supplemental Material\,\cite{PRL2023SM_Bi}). By varying the photoexcitation energies and fluences, the photo-excited carrier density and electronic temperature can be further manipulated. As an estimate, experimentally it has been reported that a fluence of $0.2-1$\,mJ/cm$^2$ leads to a thermalized electronic temperature $\Te = 5-13$\,mRy on the bismuth surface\,\cite{Faure2013}, and a $\Te > 30$\,mRy has been reported in trARPES measurement of Bi$_2$Se$_3$ with an infrared fluence of 26\,$\mu$J/cm$^2$\,\cite{Sobota2012}. The appropriate combination of $n$ and $T_e$ can be chosen from a wide region of the diagram in Fig.\,\ref{phases}(b) and (e) to monitor the structural and topological phase transitions in monolayer bismuth.


In summary, we show that light can simultaneously induce transient structural and topological phases in monolayer bismuth, allowing efficient switching between different topologies. By tailoring the photo-excited carrier density and electronic temperature, we find that four hidden crystal structures can be accessed, and in doing so the electronic and spin topology may be manipulated, which can be directly measured by existing experimental techniques that have been widely used to study photo-excited dynamics of bismuth. Our findings open a promising route towards ultrafast control of structural, electronic topological properties, as well as spin physics in emerging quantum materials.

\begin{acknowledgements}
B.P., R.J.S.~and B.M.~acknowledge funding from the Winton Programme for the Physics of Sustainability. B.P.~also acknowledges support from Magdalene College Cambridge for a Nevile Research Fellowship. G.F.L.~acknowledges funding from the Aker Scholarship. D.B.~acknowledges support from the US Army Research Office (ARO) MURI project under grant No.~W911NF-21-0147 and the National Science Foundation DMREF program under Award No.~DMR-1922172. R.J.S.~also acknowledges funding from a New Investigator Award from the EPSRC (EP/W00187X/1) and from Trinity College Cambridge. B.M.~also acknowledges support from a UKRI Future Leaders Fellowship (MR/V023926/1) and from the Gianna Angelopoulos Programme for Science, Technology, and Innovation. Calculations were performed using resources provided by the Cambridge Tier-2 system, operated by the University of Cambridge Research Computing Service (www.hpc.cam.ac.uk) and funded by EPSRC Tier-2 capital grant EP/P020259/1, as well as with computational support from the U.K. Materials and Molecular Modelling Hub, which is partially funded by EPSRC (EP/P020194), for which access is obtained via the UKCP consortium and funded by EPSRC grant EP/P022561/1.
\end{acknowledgements}

\bibliographystyle{apsrev4-1}

%

\end{document}